\documentclass{PoS}
\usepackage[square,comma,numbers,sort&compress]{natbib}

\newcommand*{\no}{\noindent}
\newcommand*{\bea}{\begin{eqnarray}}
\newcommand*{\eea}{\end{eqnarray}}
\newcommand*{\be}{\begin{equation}}
\newcommand*{\ee}{\end{equation}}

\newcommand*{\pref}[1]{(\ref{#1})}

\newcommand*{\mn}{{\mu\nu}}

\newcommand*{\prefr}[2]{(\ref{#1}-\ref{#2})} 
\newcommand*{\nn}{\nonumber}

\newcommand*{\tr}{\mathrm{tr}}

\title{Employing the perturbative definition of the Higgs mass in a non-perturbative calculation}

\ShortTitle{}

\author{\speaker{Axel Maas}\thanks{Supported by the DFG under grant number MA 3935/5-1.}\\
        Institute of Theoretical Physics, Friedrich-Schiller-University Jena, Max-Wien-Platz 1, D-07743 Jena, Germany\\
        E-mail: \email{axelmaas@web.de}}

\abstract{In perturbative calculations the masses of the Higgs, the $W$s and the $Z$ are usually determined from the pole position of the corresponding gauge-dependent propagators. In full non-perturbative lattice calculations it is much more direct to instead investigate the bound state spectrum with its gauge-independent meaning, which then contains bound states of Higgses and/or $W$s and $Z$s. It is possible to extend the perturbative definition of the Higgs mass also to such a full non-perturbative setting by determining the respective full non-perturbative propagators of the Higgs, the $W$s, and the $Z$, and analyze their analytic structure. This helps connecting the Higgs properties indirectly with gauge-invariant physics. This is here studied, using lattice gauge theory, for the case of a $W$-$Z$-Higgs system.}

\FullConference{ The XXIX International Symposium on Lattice Field Theory - Lattice 2011\\
July 10-16, 2011\\
Squaw Valley, Lake Tahoe, California}

\begin{document}

\section{The Higgs mass in and beyond perturbation theory}

The precise definition of the mass of a particle is a highly non-trivial problem, in particular in case of unstable particles. This is even more the case for (renormalizable) gauge theories. The most direct definition of the mass of a particle is just given by the (time-like) poles of its propagator \cite{Bohm:2001yx}. These poles move to complex values if the particle is unstable, in particular off the first Riemann sheet. Nonetheless, the pole mass (and width) remains a well-defined concept.

In perturbative calculations, where the analytic structures of the propagators are known, such poles can be directly identified. This becomes complicated if the propagators are not known in a closed analytic form, which is usually the case in non-perturbative calculations in Euclidean space-time. In these cases, various approaches like fitting the propagators to candidate analytic forms \cite{Maas:2011se} or approximate reconstruction of the spectral function, e.\ g.\ by use of the maximum entropy method \cite{Jarrell:1996aa}, is the approach of choice, despite the inherent systematic uncertainties \cite{Alkofer:2003jj,Nickel:2006mm}.

In a (non-)renormalizable theory, these technical problems become supplemented by conceptual ones. The masses of the elementary particles are usually no longer renormalization-group invariants, and are furthermore scheme-dependent \cite{Bohm:2001yx}. They do therefore no longer represent physical observables, but can only be determined at a given renormalization scale in a given renormalization scheme. The translation between different schemes and scales can usually be done straightforwardly in perturbation theory. Beyond perturbation theory, this is a non-trivial problem.

The situation becomes even worse when the theory at hand is a gauge theory. In this case, the elementary degrees of freedom are no longer gauge-invariant, and therefore the associated pole masses may also no longer be so. It seems that the best that can be achieved is gauge-parameter independence, i.\ e., that within a certain class of gauges the masses are gauge-invariant, as dictated for covariant gauges by Nielsen identities \cite{Nielsen:1975fs}. However, these will break down in general when moving to a different class of gauges \cite{Nielsen:1975fs}. Nevertheless, it is still possible to give a precise definition of a mass in a fixed gauge. E.\ g., in the weak sector of the standard model usually a 't Hooft gauge is chosen \cite{Bohm:2001yx}, which is a class of gauges in which the masses of the perturbatively physical degrees of freedom are not depending on the representative \cite{Bohm:2001yx}. Beyond perturbation theory, due to the Gribov-Singer ambiguity, this problem also becomes usually worse, though the latter effect may be mild for the Higgs sector of the standard model \cite{Maas:2010nc}.

As a consequence, full non-perturbative calculations usually use gauge-invariant quantities to characterize a theory. The simplest is the bound state spectrum, as is done in lattice gauge theory \cite{Gattringer:2010zz}. However, this implies that the relevant degrees of freedom are composite objects, made e.\ g.\ from two Higgs particles in a scalar isoscalar $0^{++}$ state \cite{Philipsen:1996af,Maas:2010nc}. Such an object would be rather hard to identify experimentally, in particular as it could be expected to be rather short-lived, and may not have much more structure like the notorious $\sigma$-meson of QCD. Nonetheless, in a lattice calculation, in particular when neglecting everything but the weak isospin sector and the Higgs sector, the properties of these particles can be determined \cite{Philipsen:1996af,Langguth:1985dr,Jersak:1985nf}. Similar considerations also apply for the $W$ and $Z$ bosons, with corresponding bound states, called $W$-balls and $Z$-balls \cite{Philipsen:1996af,Maas:2010nc}.

It is of course desirable to connect both descriptions. This is the aim here, where simultaneously in a full non-perturbative calculation the bound-state spectrum and the mass of the Higgs and the gauge bosons will be determined from their respective (gauge-dependent) propagators. In these preliminary results of this exploratory investigation, only the weak gauge bosons and the Higgs are included, which implies that the $W$ and the $Z$ will be degenerate, and commonly denoted as $W$.

\section{The physical spectrum}

The setup of the lattice simulation of the theory is given in \cite{Maas:2010nc}. Here, since only the approach of connecting both concepts will be investigated, a single lattice setup with fixed volume and fixed lattice spacing deep inside the Higgs phase will be used. Since only one lattice setup is used, the scale is arbitrary, and will be set to 231 GeV. The infinite-volume and continuum limit as well as the question of triviality will be addressed elsewhere \cite{Maas:unpublished}.

Since there are three free parameters in the theory, the gauge coupling, the Higgs mass, and the Higgs self-interaction, it is necessary to determine at least three different observables to fix these. These are selected to be three bound states of the $W$ and the Higgs particles.

The first is the scalar isoscalar $0^{++}$ Higgsonium \cite{Philipsen:1996af,Langguth:1985dr,Jersak:1985nf,Maas:2010nc}, a bound state of two Higgs particles, defined by the operator $H^{+}(x)H(x)$, where $H$ is the Higgs field. The second is the $W$-ball \cite{Philipsen:1996af}, defined analogously from the operator $W_\mn^a(x) W^a_\mn(x)$, where $W_\mn^a$ is the field strength of the $W$ bosons. Note that often in the lattice literature, originating from \cite{Langguth:1985dr,Jersak:1985nf}, these bound states are referred to as Higgs and $W$ instead of Higgsonium and $W$-ball, see for a discussion of this \cite{Maas:2010nc}.

\begin{figure}
\includegraphics[width=\linewidth]{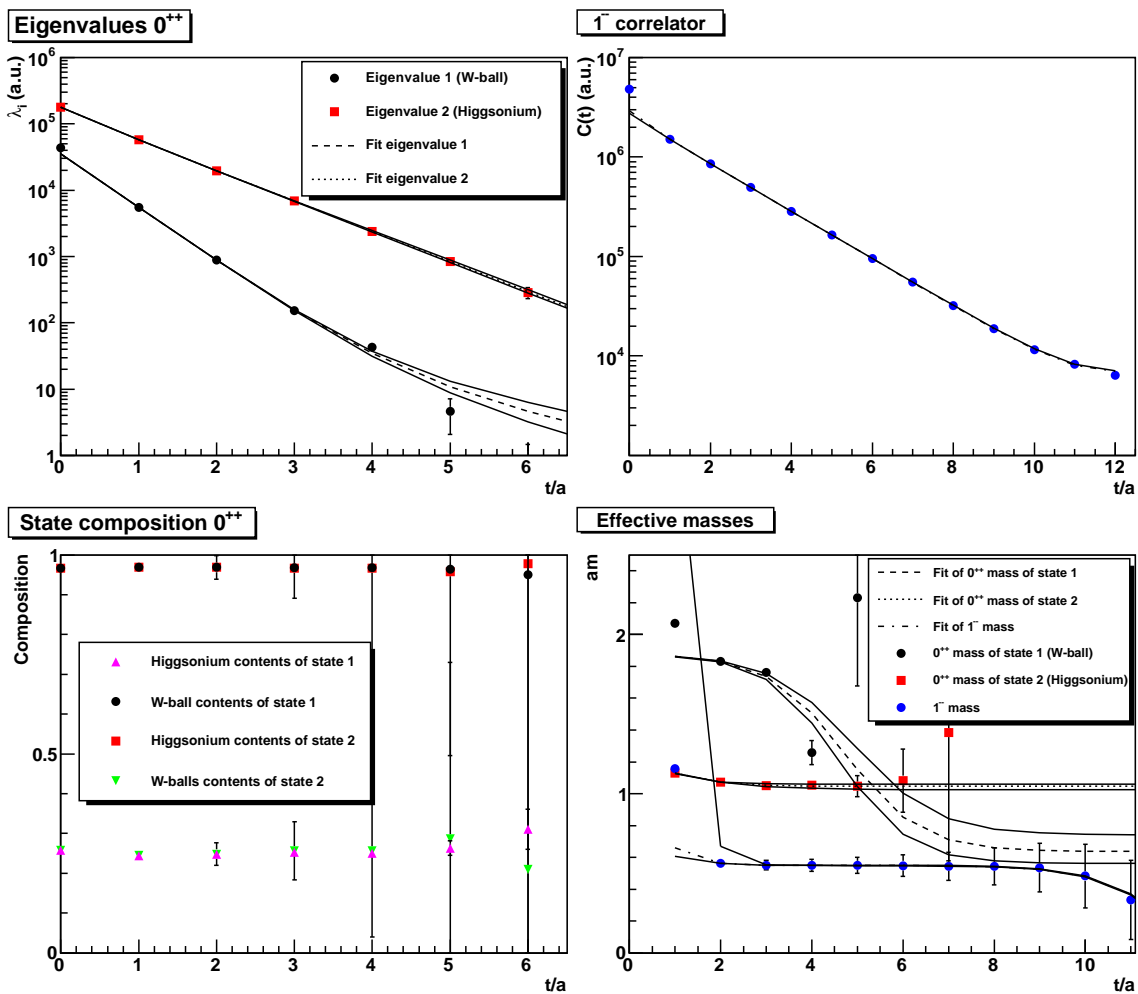}
\caption{\label{fig:spec}The low-lying spectrum. The top-left panel shows the time-dependence of the two lowest eigenvalues of the $0^{++}$ channel. The composition of the corresponding states is given in the bottom-left panel. The extracted masses are shown in the bottom-right panel. The correlation function of the lowest $1^{--}$ state is given in the top-right panel, and its mass is shown in the bottom-right panel. Errors and error bands for the fits are purely statistical at the 1$\sigma$ level. Lattice parameters are size $24^4$, $\kappa=0.32$, $\beta=2.3$, $\lambda=1$ \cite{Maas:2010nc}.}
\end{figure}

Since both operators mix with the vacuum, and thus with each other, their determination is statistically hard. To reduce the statistical fluctuations, five levels of smearing with the methods of \cite{Philipsen:1996af} are applied. It is found that at least the masses of the lowest excitations are independent, within statistical uncertainty, of the smearing process. To disentangle the two components, an eigenvalue decomposition \cite{Gattringer:2010zz} is performed. The results of these calculations can be seen in figure \ref{fig:spec}. Despite the large statistical noise, even for the employed 780000 configurations, a clear separation of both states with little mixing can be found. The light state is the Higgsonium with a mass of 242(5) GeV, while the $W$-ball has a mass of 433(3) GeV. This would correspond roughly to a constituent mass of the Higgs and the $W$ of 121(3) and 217(2) GeV, respectively.

However, such a constituent picture cannot be as simple as in QCD. In figure \ref{fig:spec} also the lowest vector isovector $1^{--}$ state is shown, which can be constructed from a combination of the Higgs and the $W$ field as \cite{Philipsen:1996af,Langguth:1985dr,Jersak:1985nf}
\bea
O^{1^{--}}_{\mu a}(x)&=&\tr \tau_a \phi^+(x) U_\mu(x)\phi(x+e_\mu)\nn\\
H&=&\rho\phi\nn,
\eea
\no where $\rho$ is the length of the Higgs field, and $\phi$ is an SU(2) matrix encoding the remainder of the Higgs degrees of freedom, $\tau^a$ is a Pauli matrix, and $U_\mu$ is a link variable of the lattice setup, behaving like the gauge field in the continuum limit. If the Higgs length fluctuates weakly around its average value, this state can be interpreted as a vector composite of at least two Higgses and a $W$. If the Higgs length fluctuates strongly, the matrix $\phi$ cannot be approximated by a finite polynomial in the Higgs field, and the structure of this bound state has to be regarded as a collective state.

The correlation function and its fitted mass is also shown in figure \ref{fig:spec}. Its mass is actually lower than that of the two $0^{++}$ states, about 127(1) GeV. Such a hierarchy has been already observed earlier \cite{Langguth:1985dr}. Here, it implies that a simple constituent model, which would imply this state to be heavier, does not apply. If this state is assumed to be made up of two collective modes, their mass would be of order 64(1) GeV.

Thus, for the present lattice setup and a constituent picture, a Higgs mass of size 121 GeV and a $W$ mass of 217 GeV would be expected, but the existence of a light state of about 64 GeV would be anticipated as well. In the next section, this will be compared with the perturbative definition of both the Higgs and the $W$ mass, using the corresponding gauge-dependent propagators.

\section{W- and Higgs-propagators and Schwinger functions}

In the following the gauge-fixed propagators of both the $W$ and the Higgs will be discussed. For this purpose, a non-aligned minimal Landau gauge \cite{Maas:2011se,Maas:unpublished} will be used, which is constructed such that the Higgs expectation value always vanishes \cite{Maas:2010nc,Maas:unpublished}. Therefore, it is not necessary to distinguish between diagonal and off-diagonal propagators, and in particular $W$ and $Z$ bosons will have identical propagators, as long as no hypercharge is introduced. The same applies to the Higgs and the would-be Goldstone bosons' propagators. This simplifies the analysis significantly. The actual lattice calculation of the propagators in this gauge is straightforward, see \cite{Maas:2010nc,Maas:unpublished} for details. 

\begin{figure}
\includegraphics[width=\linewidth]{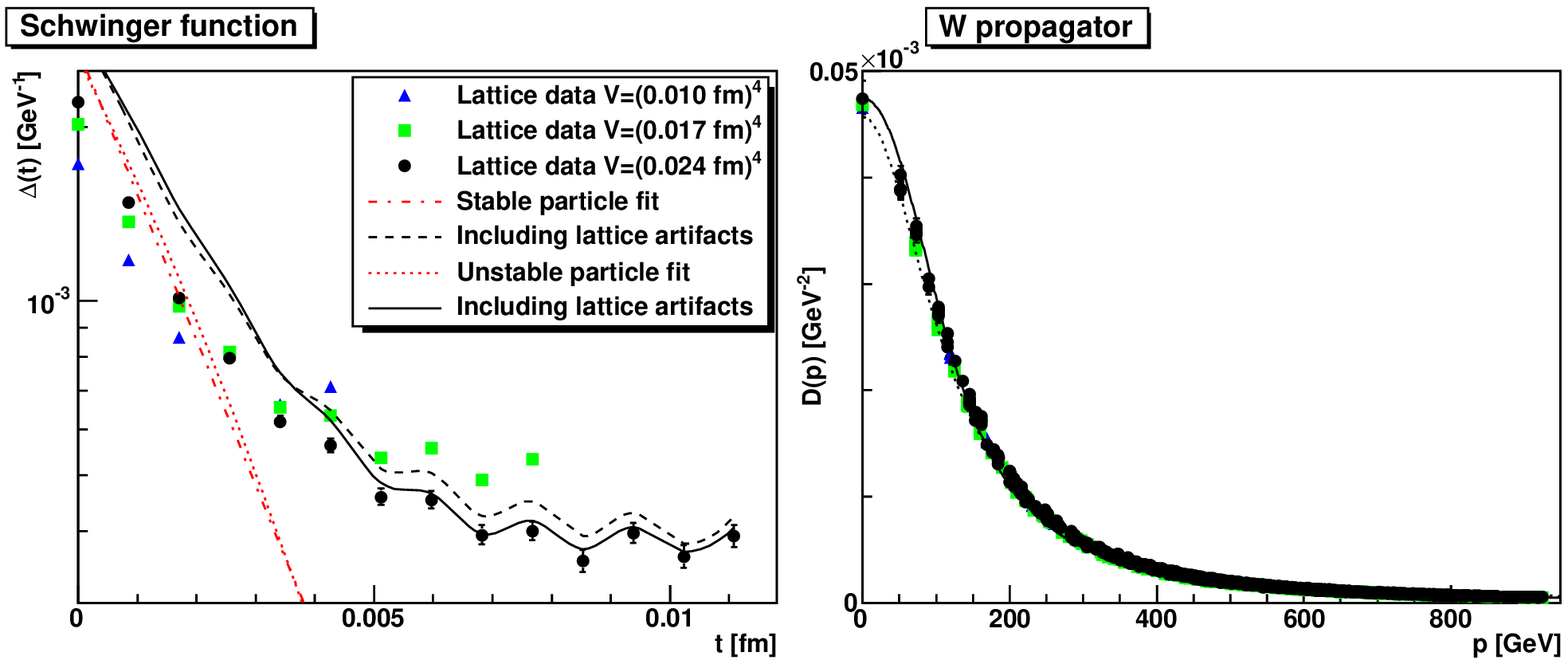}\\
\includegraphics[width=\linewidth]{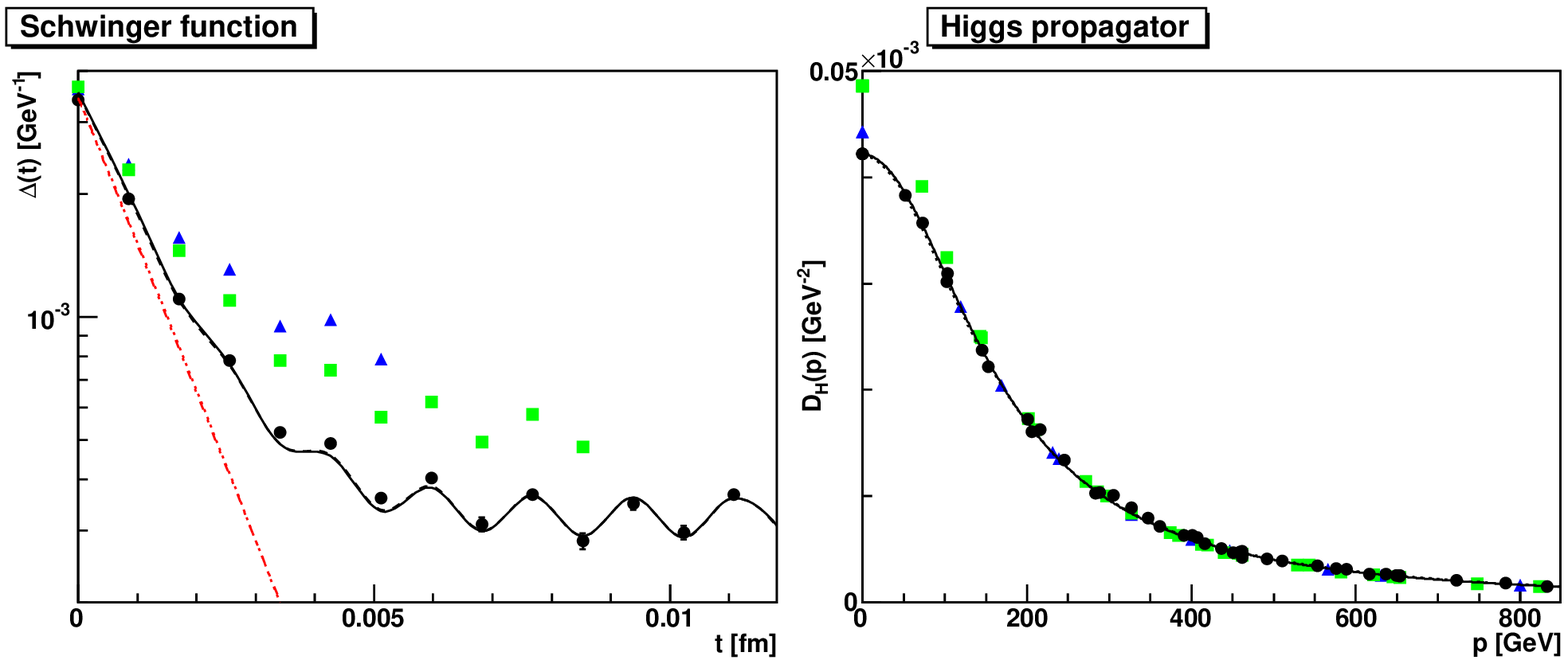}
\caption{\label{fig:props}The propagator for the $W$ (top panels) and the Higgs (bottom panels) in momentum space (right panels) and the associated Schwinger function (left panels) for different volumes. The fits shown are for the largest volumes only.}
\end{figure}

The only thing remaining is how the renormalization conditions will be chosen. This will be done here by requiring that the propagators of $W$ and the Higgs are at $\mu=250$ GeV as similar as possible to the tree-level propagators of particles with masses of 80 GeV and 125 GeV, respectively. For the $W$, this is only an overall wave-function renormalization, while for the Higgs its derivative also has to match this condition. The results, for a few different volumes, are shown in figure \ref{fig:props}. There are a number of observations.

Concerning the $W$-boson, very small finite-volume effects are seen, in stark contrast to Yang-Mills theory \cite{Maas:2011se}. The propagator itself can be fitted using an unstable\footnote{This fit has complex conjugate poles, and thus not the correct analytic structure for a genuine particle. However, the logarithmic contributions, which would move the poles to the correct Riemann sheets \cite{Bohm:2001yx} are too minor a correction than that they are detectable with the present lattice results. Thus, the approximation of complex conjugate poles is made.} particle's propagator ansatz
\be
D(p)=\frac{e^2+f p^2}{p^4+2m^2\cos\left(2\phi\right)p^2+m^4}\label{unstable},
\ee
\no though even a fit with a stable particle's propagator
\be
D(p)=\frac{1}{p^2+M^2}\label{stable},
\ee
\no gives an acceptable description. However, fits are always problematic if the exact analytic structure is unknown \cite{Alkofer:2003jj}. In general, any realistic fit must reproduce all transformations of the function to be fitted. To check this, the Schwinger function \cite{Alkofer:2003jj}, i.\ e.\ the temporal Fourier transform, is also shown in figure \ref{fig:props}. This shows that the analytic continuum transformations of \prefr{unstable}{stable} are not adequate. Including lattice artifacts by making the discrete Fourier transformation for the fits
\be
\Delta(t)=\frac{1}{a\pi}\frac{1}{N_t}\sum_i\cos\left(\frac{2\pi tp_0}{N_t}\right)D(p_0^2,\vec{0})\nn,
\ee
\no provides a much better description, though still being qualitatively incorrect at small times. A possible reason is that the mass included is too hard, and an even softer mass could be necessary.

The situation changes for the Higgs. Here, both the propagator and the Schwinger function are equally well fitted by an ansatz with a stable and an unstable particle's propagator. Furthermore, the quality of the fit for both the propagator and the Schwinger function is much better. Note that the apparent unsystematic volume dependence is rather originating from the renormalization, as the momentum spacing due to the lattice is entering sensitively in the calculation of the derivatives.

The fits permit to extract the would-be pole masses. These are listed, together with the expected constituent mass and the screening mass, i.\ e.\ $D(0)^{-1/2}$  \cite{Maas:2011se}, in table \ref{tab:masses}. It is visible how the different definitions of the mass yield different results. Note that the extracted pole masses of the $W$ are independent of the renormalization condition. The mass of the Higgs depends on the renormalization condition, due to its additive mass renormalization \cite{Bohm:2001yx}. The constituent masses are renormalization condition independent, and thus different from the actual Higgs mass.

The remaining problem is that none of these masses needs to coincide with the masses extracted from the true propagators in the complex plane \cite{Alkofer:2003jj}. Fits, even more involved ones like the maximum entropy method with its inherent problems \cite{Nickel:2006mm}, can at best indicate the correct analytic structure. Thus, eventually a full calculation in the complex plane will be required.

\begin{table}
\begin{tabular}{|c|c|c|c|c|c|c|}
 \hline
Particle & Renormalized & Screening & Stable & Unstable & Width & Constituent\cr
\hline
$W$ & 80 & 145(3) & 114(2) & 110(2) & 46(2) & 217(2) \cr
\hline
Higgs & 125 & 154(1) & 164(4) & 151(8) & 20(1) & 121(3) \cr
\hline
\end{tabular}
 \caption[]{\label{tab:masses}The various definitions of masses extracted from the propagators, as discussed in the text. ``Renormalized'' is the mass implemented as good as possible by the renormalization conditions, ``Screening'' is the screening mass, ``Stable'' is the mass extracted from a stable particle's propagator fit \protect{\pref{stable}}, ``Unstable'' and ``Width'' are the values from an unstable particle's propagator fit \protect{\pref{unstable}}, and ``Constituent'' is the constituent picture mass from the previous section. All values in GeV.}
\end{table}

\section{Conclusions}

Comparing the various definitions of the Higgs and $W$ masses given, the final value depends strongly on the way of determining it. It is furthermore not evident that any simple relation, like in constituent quark models, exists between the masses of the $W$ and the Higgs and the masses of gauge-invariant bounds. It is in fact not even clear whether the $W$ and Higgs masses are eventually gauge-invariant, it is just necessary to think of the impact of the choice of a gauge-dependent renormalization scheme. Thus, the extraction directly from the corresponding gauge-dependent propagator, like in perturbation theory, seems still to be the best defined approach.

Of course, these problems repeat themselves even if the Higgs is not elementary but composite, e.\ g.\ in technicolor theories \cite{Andersen:2011yj}. Since a composite Higgs degree of freedom behaves at low energies like in the present case, the composite state is necessarily also gauge-dependent. This leads to precisely the same problems as encountered here for an elementary Higgs. Getting this under control is therefore necessary even if the Higgs is not fundamental. Due to the strong interactions involved in these cases, the Gribov-Singer ambiguity is of similar relevance as in Yang-Mills theory \cite{Maas:2011jf}, emphasizing the need for a well defined, non-perturbative gauge-fixing \cite{Maas:2011se}.

\bibliographystyle{bibstyle}
\bibliography{bib}

\end{document}